\documentclass[a4paper, 11pt, oneside]{article}
\pdfoutput=1
\usepackage{float}
\usepackage{jheppub}
\usepackage{epstopdf}

\usepackage{bbm}   
\usepackage{amsmath}
\usepackage{graphicx}

\newcommand{\nc}{\newcommand}

\newlength{\absize}
\nc{\non}{\nonumber}
\nc{\hc}{\hbox {H.c.}} 
\nc{\noi}{\noindent}
\nc{\barx}{\bar{x}}
\nc{\pbarn}{\;\hbox {pb}}
\nc{\fbarn}{\;\hbox {fb}}
\newcommand{\bi}{\begin{itemize}}
\newcommand{\ei}{\end{itemize}}

\nc{\lsp}{\;\;\;\;\;}
\nc{\Lsp}{\;\;\;\;\;\;\;\;\;\;}  
\nc{\LLsp}{\lspace \lspace}
\nc{\lra}{\longrightarrow}
\nc{\beq}{\begin{equation}}  \nc{\eeq}{\end{equation}}
\nc{\bea}{\begin{eqnarray}}  \nc{\eea}{\end{eqnarray}}
\nc{\baa}{\begin{array}}     \nc{\eaa}{\end{array}}
\nc{\bit}{\begin{itemize}}   \nc{\eit}{\end{itemize}}
\nc{\ben}{\begin{enumerate}} \nc{\een}{\end{enumerate}}
\nc{\bce}{\begin{center}}    \nc{\ece}{\end{center}}
\nc{\bpm}{\begin{pmatrix}}   \nc{\epm}{\end{pmatrix}}
\nc{\bvt}{\begin{verbatim}}  \nc{\evt}{\end{verbatim}}
\def\lsim{\mathrel{\raise.3ex\hbox{$<$\kern-.75em\lower1ex\hbox{$\sim$}}}}
\def\gsim{\mathrel{\raise.3ex\hbox{$>$\kern-.75em\lower1ex\hbox{$\sim$}}}}

\def\pcal{{\cal P}}

\def\zcal{{\cal Z}}

\def\gev{\;\hbox{GeV}}

\nc{\tanb}{\tan\beta}
\nc{\mch}{M_{H^\pm}}

\def\cb{c_\beta}
\def\sb{s_\beta}

\def\mch{M_{H^\pm}}

\nc{\for}{\lsp {\rm for} \lsp}
\nc{\andd}{\lsp {\rm and} \lsp}

\renewcommand{\Re}{\mbox{Re\thinspace}}
\renewcommand{\Im}{\mbox{Im\thinspace}}

\allowdisplaybreaks 

\def\i11{{\mathbbm 1}}

\title{Spontaneous CP violation in the 2HDM: \\ physical conditions and the alignment limit}
\author[a]{B. Grzadkowski,}
\affiliation[a]{Faculty of Physics, University of Warsaw, Pasteura 5, 02-093 Warsaw, Poland}
\author[b]{O. M. Ogreid,}
\affiliation[b]{Bergen University College, Postboks 7030, N-5020 Bergen, Norway}
\author[c]{P. Osland}
\affiliation[c]{Department of Physics and Technology,
University of Bergen, Postboks 7803, N-5020 Bergen, Norway}
\emailAdd{bohdan.grzadkowski@fuw.edu.pl}
\emailAdd{omo@hib.no}
\emailAdd{Per.Osland@ift.uib.no}
\date{\today}

\abstract{For the general Two-Higgs-Doublet model, we present
  conditions for having spontaneous CP violation, in terms of physical
  masses and couplings. These relations involve the charged-Higgs mass,
	its cubic couplings with neutral scalars and quartic coupling, and become particularly simple in the
  alignment limit. In the simplified model with softly broken $Z_2$
  symmetry, some deviation from alignment is required for spontaneous CP violation to be present.}


\begin{document}

\maketitle

\flushbottom

\section{Introduction}
\label{Sec:Introduction}
The conditions for CP violation in the general\footnote{In the {\it
    most} general 2HDM, one can also assume a charge-breaking vacuum
  by assigning a non-zero vacuum expectation value to the upper
  components of the Higgs doublets. We do not consider such
  models. Henceforth, when we refer to the general 2HDM, we mean the
  most general charge-conserving 2HDM.} Two-Higgs-Doublet Model (2HDM)
can be formulated in terms of physical quantities, masses of the
three neutral Higgs bosons ($M_j$), their couplings to gauge bosons
($e_j$) and their couplings to charged  Higgs bosons ($q_j$) \cite{Lavoura:1994fv,Botella:1994cs} (see also Ref.~\cite{Grzadkowski:2014ada}).
An explicitly CP conserving potential is a necessary requirement for having {\it spontaneous} CP violation (SCPV).
In order to distinguish whether the CP violation is explicit or
spontaneous, one can evaluate certain invariants (with respect to
Higgs basis transformations) formed from potential parameters
\cite{Branco:2005em,Gunion:2005ja}. 
In this paper, we show how these
conditions can be expressed in terms of physical quantities. In fact,
we find that when CP is violated, two more conditions need to be satisfied in order to guarantee that the CP violation is spontaneous: the charged Higgs boson mass and the quartic
coupling among charged Higgs bosons must be related to the masses of
the neutral Higgs bosons, and the above trilinear couplings ($e_j,\ q_j$). These relations are presented in simple forms. 

We also investigate SCPV in the alignment limit \cite{Asner:2013psa,Craig:2013hca,Carena:2013ooa}, as well as in the simplified model with softly broken $\zcal_2$ symmetry. 

The paper is structured as follows. 
In section~\ref{Sec:model} we present our notation, and quote some basic properties of the model.
In section~\ref{Sec:SCPV-general} we present our main result, a theorem specifying (in physical terms) the conditions for spontaneous CP violation. The theorem is established by expanding the $I$ invariants of Ref.~\cite{Gunion:2005ja} in terms of the $\Im J_i$ invariants (see Eq.~(\ref{Eq:Im-J}) below) and systematically setting these $I$ invariants to zero while retaining at least one $\Im J_i$ non-zero.
Next, in section~\ref{Sec:general} we discuss spontaneous CP violation in the ``Alignment'' limit, staying within the general model without $\zcal_2$ symmetry
(this model will be referred to as ``2HDM7''). Then, in section~\ref{Sec:SCPV5} we discuss the case of softly broken $\zcal_2$ symmetry, denoted ``2HDM5'', where CP violation requires some deviation from the alignment limit.
Finally, section~\ref{Sect:conclusion} contains concluding remarks.
Technical details are given in appendices A (the Higgs basis), B (explicit results for the $I$ invariants in terms of physical quantities) and C (solving the constraints $I{...}=0$).
\section{The model}
\label{Sec:model}

The scalar potential of the 2HDM might be parametrized in the standard fashion\footnote{A comment is here in order. In fact, $m_{12}^2$ is a redundant parameter which could, in the case of the general 2HDM, be removed from the potential by a unitary rotation of the doublets. Therefore, without loss of generality we could, from the very beginning, have dropped the $m_{12}^2 \Phi_1^\dagger \Phi_2 + \hc$ term. Relations among physical quantities (observables) can not be sensitive to this choice of basis. However, in order to stay within this most general basis, we have decided to keep it.}:
\begin{align}
\label{Eq:pot}
V(\Phi_1,\Phi_2) &= -\frac12\left\{m_{11}^2\Phi_1^\dagger\Phi_1
+ m_{22}^2\Phi_2^\dagger\Phi_2 + \left[m_{12}^2 \Phi_1^\dagger \Phi_2
+ \hc\right]\right\} \nonumber \\
& + \frac{\lambda_1}{2}(\Phi_1^\dagger\Phi_1)^2
+ \frac{\lambda_2}{2}(\Phi_2^\dagger\Phi_2)^2
+ \lambda_3(\Phi_1^\dagger\Phi_1)(\Phi_2^\dagger\Phi_2) 
+ \lambda_4(\Phi_1^\dagger\Phi_2)(\Phi_2^\dagger\Phi_1)\nonumber \\
&+ \frac12\left[\lambda_5(\Phi_1^\dagger\Phi_2)^2 + \hc\right]
+\left\{\left[\lambda_6(\Phi_1^\dagger\Phi_1)+\lambda_7
(\Phi_2^\dagger\Phi_2)\right](\Phi_1^\dagger\Phi_2)
+{\rm \hc}\right\} \\
&\equiv Y_{a\bar{b}}\Phi_{\bar{a}}^\dagger\Phi_b+\frac{1}{2}Z_{a\bar{b}c\bar{d}}(\Phi_{\bar{a}}^\dagger\Phi_b)(\Phi_{\bar{c}}^\dagger\Phi_d).
\label{Eq:pot-other}
\end{align}
In the second form \cite{Branco:1999fs}, Eq.~(\ref{Eq:pot-other}), a summation over barred with un-barred indices is implied, e.g., $a=\bar a=1,2$.  For the explicit expressions of the $Y_{a\bar{b}}$ and $Z_{a\bar{b}c\bar{d}}$ tensors, see for example Eqs.~(2.3) and (2.4) in \cite{Grzadkowski:2014ada}.

In general the vacuum may be complex, and the Higgs doublets can be parametrized as 
\begin{equation}
	\Phi_j=e^{i\xi_j}\left(
	\begin{array}{c}\varphi_j^+\\ (v_j+\eta_j+i\chi_j)/\sqrt{2}
	\end{array}\right), \quad
	j=1,2.\label{vevs}
\end{equation}
Here, $v_j$ are non-negative real numbers, so that $v_1^2+v_2^2=v^2=(246 \gev)^2$. The fields $\eta_j$ and $\chi_j$ are real.
The phase difference between the two vevs will be denoted as $\xi\equiv\xi_2-\xi_1$.

We also follow standard terminology and define\footnote{This parameter is only meaningful in a basis where both $v_i>0$, so in particular, for the Higgs-basis it is not meaningful.}
\begin{equation} \label{Eq:musq}
\mu^2\equiv\frac{v^2\Re m_{12}^2}{2v_1v_2}.
\end{equation}

The model contains three neutral scalars $H_i$. The mass eigenstates can be obtained by an orthogonal rotation upon the $\eta_i$ as follows:
\begin{equation} \label{Eq:R-def}
\begin{pmatrix}
H_1 \\ H_2 \\ H_3
\end{pmatrix}
=R
\begin{pmatrix}
\eta_1 \\ \eta_2 \\ \eta_3
\end{pmatrix},
\end{equation}
where the $3\times3$ orthogonal rotation matrix $R$ satisfies
\begin{equation}
\label{Eq:cal-M}
R{\cal M}^2R^{\rm T}={\cal M}^2_{\rm diag}={\rm diag}(M_1^2,M_2^2,M_3^2),
\end{equation}
with $M_1\leq M_2\leq M_3$. A convenient parametrization of the rotation matrix $R$ is \cite{Accomando:2006ga,ElKaffas:2006gdt}
\begin{equation} \label{Eq:R_alpha}
R=
\begin{pmatrix}
R_{11}    &  R_{12}   & R_{13}   \\
R_{21}    &  R_{22}   & R_{23}   \\
R_{31}    &  R_{32}   & R_{33}  
\end{pmatrix}
=
\begin{pmatrix}
c_1\,c_2 & s_1\,c_2 & s_2 \\
- (c_1\,s_2\,s_3 + s_1\,c_3) 
& c_1\,c_3 - s_1\,s_2\,s_3 & c_2\,s_3 \\
- c_1\,s_2\,c_3 + s_1\,s_3 
& - (c_1\,s_3 + s_1\,s_2\,c_3) & c_2\,c_3
\end{pmatrix},
\end{equation}
where $c_i=\cos\alpha_i$ and $s_i=\sin\alpha_i$.

Much of our discussion will be phrased in terms of the four physical masses of the model,
\begin{equation}
M_1 \leq M_2 \leq M_3, \quad \text{and}\quad M_{H^\pm},
\end{equation}
together with parameters $e_i$, $q_i$ and $q$.
The $e_i$ are defined as
\begin{equation}
e_i \equiv v_1R_{i1}+v_2R_{i2}\label{eq:e_i-def}.
\end{equation}
They parametrize the couplings of neutral Higgs bosons to the gauge particles $Z$ and $W$, whereas the other couplings \cite{Grzadkowski:2014ada},
\begin{align}
q_{i}&\equiv\text{Coefficient}(V,H_iH^-H^+)\label{eq:qi}
\nonumber\\
&=\frac{2 e_i}{v^2}M_{H^\pm}^2
-\frac{R_{i2} v_1+R_{i1} v_2}{v_1 v_2}\mu^2
+\frac{g_i}{v^2 v_1 v_2}M_i^2
+\frac{R_{i3} v^3}{2 v_1 v_2}\Im\lambda_5\nonumber\\
&+\frac{v^2 \left(R_{i2} v_1-R_{i1} v_2\right)}{2 v_2^2}\Re\lambda_6
-\frac{v^2 \left(R_{i2} v_1-R_{i1} v_2\right)}{2 v_1^2}\Re\lambda_7, \\
q&\equiv\text{Coefficient}(V,H^-H^-H^+H^+) \\
	&=-\frac{1}{2 v^2 v_1^2 v_2^2}
	\left(v_1^2-v_2^2\right)^2\mu^2
	+\sum_{k=1}^3\frac{g_k^2}{2 v^4 v_1^2 v_2^2}M_k^2 
	+\frac{v^2\left(
		v_1^2-3 v_2^2\right)}{4 v_1 v_2^3} \Re\lambda_6
	+\frac{v^2\left(v_2^2-3 v_1^2\right)}{4 v_2 v_1^3}\Re\lambda_7\label{Eq:coupling-q}
	\nonumber
\end{align}
describe the trilinear couplings ($q_i$) of neutral Higgs bosons to the charged pair, as well as the quartic coupling ($q$) among four charged Higgs bosons, respectively.
Here, we have chosen a basis for which $\xi=0$. We shall continue to use such a basis, unless otherwise stated.
Also, we define the abbreviation
\begin{equation}
g_j\equiv v_1^3R_{j2}+v_2^3R_{j1}.
\end{equation}

\section{Spontaneous CP violation in the general case}
\label{Sec:SCPV-general}

The presence of CP violation in the Higgs sector can be expressed in terms of three invariants, at least one of which should be non-zero \cite{Gunion:2005ja} for CP violation to occur:
\begin{subequations} \label{Eq:Im-J}
\begin{align} \label{eq:im_J1}
\Im J_1&=-\frac{2}{v^2}\Im\bigl[\hat{v}_{\bar{a}}^* Y_{a\bar{b}} Z_{b\bar{d}}^{(1)}\hat{v}_d\bigr], \\
\label{eq:im_J2}
\Im J_2&=\frac{4}{v^4}\Im\bigl[\hat{v}_{\bar{b}}^* \hat{v}_{\bar{c}}^* Y_{b\bar{e}} Y_{c\bar{f}} Z_{e\bar{a}f\bar{d}}\hat{v}_a\hat{v}_d\bigr], \\
\Im J_3&=\Im\bigl[\hat{v}_{\bar{b}}^* \hat{v}_{\bar{c}}^* Z_{b\bar{e}}^{(1)} Z_{c\bar{f}}^{(1)}Z_{e\bar{a}f\bar{d}}\hat{v}_a\hat{v}_d\bigr]\,.
 \label{eq:im_J3}
\end{align}
\end{subequations}
In \cite{Grzadkowski:2014ada}, all the $\Im J_i$ were expressed in terms of masses and couplings, there we also replaced $\Im J_3$ by a related, slightly modified invariant which we named $\Im J_{30}$. The condition for CP violation was reformulated so that at least one of the invariants $\Im J_1, \Im J_2, \Im J_{30}$ should be non-zero for CP violation to occur. Expressed in terms of masses and couplings they become \cite{Lavoura:1994fv,Botella:1994cs,Grzadkowski:2014ada}
\begin{subequations} \label{Eq:Im-J-physical}
\begin{eqnarray}
\Im J_1&=&\frac{1}{v^5}\sum_{i,j,k}\epsilon_{ijk}M_i^2e_ie_kq_j\nonumber\\
&=&\frac{1}{v^5}[e_1e_2q_3(M_2^2-M_1^2)-e_1e_3q_2(M_3^2-M_1^2)+e_2e_3q_1(M_3^2-M_2^2)],\label{eq:imjJ-physical}\\
\Im J_2&=&\frac{2}{v^9}\sum_{i,j,k}\epsilon_{ijk}e_ie_je_kM_i^4M_k^2=\frac{2e_1 e_2 e_3}{v^9}\sum_{i,j,k}\epsilon_{ijk}M_i^4M_k^2\nonumber\\
&=&\frac{2e_1 e_2 e_3}{v^9}(M_2^2-M_1^2)(M_3^2-M_2^2)(M_3^2-M_1^2),\label{eq:imJ2-physical}\\
\Im J_{30}&\equiv&\frac{1}{v^5}\sum_{i,j,k}\epsilon_{ijk} q_i M_i^2  e_j q_k,\nonumber\\
&=&\frac{1}{v^5}[q_1q_2e_3(M_2^2-M_1^2)-q_1q_3e_2(M_3^2-M_1^2)+q_2q_3e_1(M_3^2-M_2^2)]\label{eq:imJ3-physical}.
\end{eqnarray}
\end{subequations}
Another related quantity that we shall encounter is $\Im J_{11}$. This quantity is given as
\bea
\Im J_{11}&\equiv&\frac{1}{v^7}\sum_{i,j,k}\epsilon_{ijk}e_i M_i^2 M_j^2 e_k q_j \\
&=&\frac{1}{v^7}[e_1e_2q_3M_3^2(M_2^2-M_1^2)-e_1e_3q_2M_2^2(M_3^2-M_1^2)+e_2e_3q_1M_1^2(M_3^2-M_2^2)].\nonumber
\eea
This quantity also vanishes whenever we have CP conservation, more specifically $\Im J_{11}$ vanishes when both $\Im J_1$ and $\Im J_2$ vanish.

In order for CP violation to be spontaneous, at least one of the $\Im J_i$ invariants must be non-zero, while four other weak-basis invariants (hereafter referred to as the $I$-invariants), constructed from the coefficients of the potential, must vanish \cite{Branco:2005em}.
These can be expressed as \cite{Davidson:2005cw,Gunion:2005ja}:
\begin{subequations} \label{Eq:I}
\begin{align}
I_{Y3Z}&=\Im \bigl[Z_{a\bar{c}}^{(1)} Z_{e\bar{b}}^{(1)} Z_{b\bar{e}c\bar{d}}Y_{d\bar{a}}\bigr],\\
I_{2Y2Z}&=\Im \bigl[Y_{a\bar{b}}  Y_{c\bar{d}} Z_{b\bar{a}d\bar{f}} Z_{f\bar{c}}^{(1)}\bigr],\\
I_{3Y3Z}&=\Im \bigl[Z_{a\bar{c}b\bar{d}} Z_{c\bar{e}d\bar{g}} Z_{e\bar{h}f\bar{q}} Y_{g\bar{a}} Y_{h\bar{b}} Y_{q\bar{f}}\bigr],\\
I_{6Z}&=\Im \bigl[Z_{a\bar{b}c\bar{d}}Z_{b\bar{f}}^{(1)} Z_{d\bar{h}}^{(1)} Z_{f\bar{a}j\bar{k}}Z_{k\bar{j}m\bar{n}}Z_{n\bar{m}h\bar{c}}\bigr].
\end{align}
\end{subequations}
Although helpful to determine the CP nature of a particular model, these expressions offer little physical insight when written out in terms of the parameters of the potential\footnote{An alternative formulation of these conditions has been given in terms of coefficients of bilinears in the fields \cite{Maniatis:2007vn}.}. 

A major result of this paper is the re-expression of these invariants in terms of masses and couplings in a similar way as was done for the $\Im J_i$ invariants in \cite{Grzadkowski:2014ada}. The results can be found in Appendix~\ref{sect:results}, and conditions for their vanishing are discussed in Appendix~\ref{sect:newresults}. As pointed out by Gunion and Haber \cite{Gunion:2005ja}, at most two invariants need to be considered for any given model, see the discussion in Sec.~III of \cite{Gunion:2005ja} for details. From the results in Appendix~\ref{sect:newresults}
we conclude that eqs.~(\ref{Eq:Mh_ch_condition}) and (\ref{Eq:q_condition}) cover all cases.
This enables us to formulate the conditions for spontaneous CP violation in a compact and elegant way:

\paragraph{Theorem.}
Let us assume that the quantity 
\bea
D=e_1^2M_2^2M_3^2+e_2^2M_3^2M_1^2+e_3^2 M_1^2 M_2^2
\eea
is non-zero\footnote{In any physical model we demand that all $M_i^2>0$ and that at least one $e_i$ is non-zero, making $D$ a positive definite quantity.}. Then,
in a charge-conserving general 2HDM, CP is violated spontaneously if and only if the following three statements are satisfied simultaneously:
\begin{alignat}{2}
	&\!\!\bullet &\ \
	&\text{At least one of the three invariants $\Im J_1$, $\Im J_2$, $\Im J_{30}$ is nonzero.} \nonumber \\
	&\!\!\bullet &\ \
	&M_{H^\pm}^2=\frac{v^2}{2D}\label{eq:theorem-mch}
	[e_1q_1M_2^2M_3^2 + e_2q_2 M_3^2 M_1^2 + e_3 q_3 M_1^2 M_2^2-M_1^2 M_2^2 M_3^2], \\
	&\!\!\bullet &\ \
	&q=\frac{1}{2D}\label{eq:theorem-q}
	[(e_2q_3-e_3q_2)^2M_1^2+(e_3q_1-e_1q_3)^2M_2^2+(e_1q_2-e_2q_1)^2M_3^2 + M_1^2 M_2^2 M_3^2].
\end{alignat}

We note that Eq.~(\ref{Eq:I}) represents {\it four} conditions, while here we only have {\it two.} This could be understood as follows. The general potential contains {\it four} complex parameters, $m_{12}^2$, $\lambda_5$, $\lambda_6$ and $\lambda_7$. However, it is always possible to remove $m_{12}^2$ by a basis rotation. Then, one may make either of $\lambda_5$, $\lambda_6$ or $\lambda_7$ real by a phase rotation of $\Phi_1$ relative to $\Phi_2$. Thus, we are down to {\it two} complex phases, matching the fact that there are only {\it two} conditions. As discussed
in \cite{Gunion:2005ja} the {\it four} conditions are needed to cover all possible models one can construct.
\section{SCPV and alignment within the general model (``2HDM7'')}
\label{Sec:general}
Our goal hearafter is to discuss a minimal setup which allows for $H_1$ to be identified with the observed $125 \gev$ state
{\it together} with spontaneous CP violation. 
As we have shown earlier \cite{Grzadkowski:2014ada} CP violation in
the exact alignment limit requires the general 2HDM with no $\zcal_2$ symmetry imposed. 

\subsection{Alignment}
Here, we define alignment in four superficially different, but equivalent ways. The most physical definition is in terms of the gauge-Higgs couplings $e_i$.
As we recall next, the alignment condition can also be formulated in terms of quartic couplings of the potential and vevs. Next, using the minimization conditions, the relations among quartic couplings can be re-expressed as conditions on the bilinear terms. Finally, it can be expressed in terms of elements of the rotation matrix in the neutral sector.
For completeness we also show the relevant couplings $q_i$ and $q$ in the alignment limit.

\subsubsection{Alignment defined in terms of physical couplings}

We define the alignment limit as the limit in which the discovered Higgs boson (assumed to be $H_1$) has Standard-Model coupling to the gauge bosons. In our terminology, this means
\begin{equation}
e_1=v, \quad e_2=e_3=0.
\label{ali_def}
\end{equation}
Since these couplings satisfy $e_1^2+e_2^2+e_3^2=v^2$, these are {\it two} conditions.
In this limit, only one of the above-mentioned $\Im J$ invariants is non-zero:
\begin{subequations}
\begin{align}
\Im J_1 &=\Im J_2 =0, \\
\Im J_{30} &=\frac{q_2q_3}{v^4}(M_3^2-M_2^2).
\end{align}
\end{subequations}
Thus, in order to have any CP-violation at all in this limit, we must have
\begin{equation}
q_2\neq0, \quad 
q_3\neq0, \quad
M_2\neq M_3.
\end{equation}
The two heavier neutral Higgs bosons must have different masses, and they must both have non-vanishing couplings to the charged pair. The non-degeneracy of $M_2$ and $M_3$ requires \cite{Khater:2003wq,Grzadkowski:2014ada}
\begin{equation}
\Re\lambda_6\neq0, \quad\text{and/or} \quad
\Re\lambda_7\neq0.
\end{equation}

\subsubsection{Alignment defined in terms of quartic potential couplings and vevs}
The alignment conditions can alternatively be phrased as
follows~\cite{Grzadkowski:2014ada} (for the CP conserving case see also \cite{Dev:2014yca}):
\begin{align}
	&v_1v_2\Im\lambda_5+v_1^2\Im\lambda_6+v_2^2\Im\lambda_7=0, \label{align_lam_1}\\
	&v_1^3v_2(\lambda_1-\lambda_{345})-v_1v_2^3(\lambda_2-\lambda_{345})
	-v_1^4\Re\lambda_6+v_2^4\Re\lambda_7
	+3v_1^2v_2^2\Re(\lambda_6-\lambda_7)=0, \label{align_lam_2}
\end{align}
with $\lambda_{345}\equiv\lambda_3+\lambda_4+\Re\lambda_5$.

\subsubsection{Alignment defined in terms of bilinear potential couplings and vevs}
Using the minimization conditions (A.1) and (A.2) of Ref.~\cite{Grzadkowski:2014ada}, the alignment conditions
(\ref{align_lam_1}) and (\ref{align_lam_2}) can also be formulated as
\bea
\Im m_{12}^2&=&0\label{eq:newalign1},\\
m_{11}^2-m_{22}^2&=&\Re m_{12}^2\left(\frac{v_1}{v_2}-\frac{v_2}{v_1}\right)\label{eq:newalign2}.
\eea

\subsubsection{Alignment defined in terms of the rotation matrix \boldmath{$R$} and vevs}
In terms of the rotation matrix of (\ref{Eq:R-def}) and (\ref{Eq:R_alpha}), alignment corresponds to
\begin{equation}
	\frac{R_{12}}{R_{11}}=\frac{v_2}{v_1}, \quad R_{13}=0,
\end{equation}
or
\begin{equation} \label{Eq:alignment-angles}
	\alpha_1=\beta, \quad \alpha_2=0,
\end{equation}
where $\tan\beta\equiv v_2/v_1$.
We make note of the fact that $\alpha_1$, $\alpha_2$, $v_1$ and $v_2$ are parameters of the semi-physical parameter set $\pcal_{67}$ that was defined in Eq.~(3.2) of \cite{Grzadkowski:2014ada}, so this definition of alignment immediately gives us the definition of alignment in terms of the parameter set $\pcal_{67}$,
\begin{equation} \label{Eq:alignment-p67}
	\tan\alpha_1=\frac{v_2}{v_1}, \quad \alpha_2=0.
\end{equation}

\subsubsection{Couplings in the alignment limit}
The couplings $q_i$ and $q$ simplify in the alignment limit:
\begin{subequations}\label{couplings-alignment}
\begin{align}
	q_{1}
	&=\frac{1}{v}\left(2 M_{H^\pm}^2-2\mu^2+M_1^2\right),
	\label{q1}\\
	q_{2}
	&=
	+c_3\left[\frac{(\cb^2-\sb^2)}{v \cb \sb}(M_2^2-\mu^2)
	+\frac{v}{2 \sb^2 }\Re\lambda_6
	-\frac{v}{2 \cb^2}\Re\lambda_7\right]
	+s_3\frac{v}{2 \cb \sb}\Im\lambda_5,
	\label{q2}\\
	q_{3}
	&=
	-s_3\left[\frac{(\cb^2-\sb^2)}{v \cb \sb}(M_3^2-\mu^2)
	+\frac{v}{2 \sb^2}\Re\lambda_6
	-\frac{v}{2 \cb^2}\Re\lambda_7\right]
	+c_3\frac{v}{2 \cb \sb}\Im\lambda_5.
	\label{q3}\\
	q&=-\frac{1}{2 v^2 c_\beta^2 s_\beta^2}
	\left(c_\beta^2-s_\beta^2\right)^2\mu^2
	+\frac{1}{2v^2}M_1^2
	+\frac{(c_\beta^2-s_\beta^2)^2}{2c_\beta^2s_\beta^2v^2}c_3^2M_2^2
	+\frac{(c_\beta^2-s_\beta^2)^2}{2c_\beta^2s_\beta^2v^2}s_3^2M_3^2\nonumber\\
	&+\frac{\left(
		c_\beta^2-3 s_\beta^2\right)}{4 c_\beta s_\beta^3} \Re\lambda_6
	+\frac{\left(s_\beta^2-3 c_\beta^2\right)}{4 s_\beta c_\beta^3}\Re\lambda_7.
\end{align}
\end{subequations}

\subsection{Spontaneous CP violation in the alignment limit}
\label{Sec:SCPV-alignment}
A necessary condition for having spontaneous CP-violation is that all four $I$-invariants must vanish. This guarantees CP invariance of the potential. Together with the non-vanishing of at least one $\Im J_i$, this constitutes necessary and sufficient conditions for the presence of spontaneous CP violation. 
We will in the remainder of this section implicitly assume that CP is violated ($\Im J_{30}\neq0$), and the vanishing of the four $I$-invariants may then be substituted for 
by the conditions (\ref{eq:theorem-mch}) and (\ref{eq:theorem-q}). Under the assumption of CP violation, CP invariance of the potential and spontaneous CP violation bear the same meaning.
\subsubsection{Spontaneous CP violation in the alignment limit in terms of physical couplings}
\label{Sec:SCPV-alignment-physical}
We may insert the conditions for alignment, $e_1=v, e_2=e_3=0$ into (\ref{eq:theorem-mch}) and (\ref{eq:theorem-q}), or simply inspect (\ref{Eq:Mh_ch_condition-case1}) and (\ref{Eq:q_condition-case1}):
\bea
M_{H^\pm}^2&=&\frac{vq_1-M_1^2}{2},\label{Eq:expression-couplings-alignment-one}\\
q&=&\frac{1}{2}\left(\frac{q_2^2}{M_2^2}+\frac{q_3^2}{M_3^2}+\frac{M_1^2}{v^2}\right).\label{Eq:expression-couplings-alignment-two}
\eea
These two simple equations constitute the conditions for having SCPV in the alignment limit.

By combining and rewriting these two equations, one obtains
\bea 
&&q_1=\frac{1}{v}\left(2M_{H^\pm}^2+M_1^2  \right)\label{Eq:expression-couplings-alignment-1}\\
&&q-\frac{1}{2}\left(
\frac{q_1^2}{M_1^2}+\frac{q_2^2}{M_2^2}+\frac{q_3^2}{M_3^2}\right)
=\frac{-2M_{H^\pm}^2(M_1^2+M_{H^\pm}^2)}{v^2M_1^2}.\label{Eq:expression-couplings-alignment-2}
\eea
\subsubsection{Spontaneous CP violation in the alignment limit in terms of semi-physical parameter set \boldmath{${\cal P}_{67}$}} 
\label{Sec:SCPV-alignment-p67}

Inserting the couplings of (\ref{couplings-alignment}) into the conditions for SCPV, (\ref{Eq:expression-couplings-alignment-1}) and (\ref{Eq:expression-couplings-alignment-2}), we find that the two conditions in the alignment limit translate to
\bea 
&&\mu^2=0\label{Eq:muzero},
\eea
and
\bea
&&-v^2\left[v_1^2\Re\lambda_6-v_2^2\Re\lambda_7\right]^2(M_2^2s_3^2+M_3^2c_3^2)
-v^2v_1^2v_2^2(\Im \lambda_5)^2(M_2^2c_3^2+M_3^2s_3^2)\nonumber \\
&&+2v^2v_1v_2s_3c_3(M_2^2-M_3^2)\Im\lambda_5(v_1^2\Re\lambda_6-v_2^2\Re\lambda_7)\nonumber\\
&&
-2v_1v_2M_2^2M_3^2(v_1^2\Re\lambda_6+v_2^2\Re\lambda_7)=0.\label{Eq:complicated}
\eea
The first of these two conditions refers to the real part of the bilinear coupling between $\Phi_1$ and $\Phi_2$, therefore
we conclude that alignment and SCPV together imply that this mass mixing parameter must be either pure imaginary or it must vanish altogether.

While the constraints (\ref{Eq:expression-couplings-alignment-one}) and (\ref{Eq:expression-couplings-alignment-two}) relate masses and couplings, the constraints in (\ref{Eq:muzero}) and (\ref{Eq:complicated}) are the equivalent of these, but relate parameters of the model from the parameter set ${\cal P}_{67}$.

\subsubsection{Spontaneous CP violation in the alignment limit in terms of potential parameters and vevs}
\label{Sec:SCPV-alignment-potentialparameters}
The condition (\ref{Eq:muzero}) simply translates to
\bea \label{Eq:m12sqre}
\Re m_{12}^2&=&0.
\eea
Using the relation between the mass matrix elements, the rotation matrix and the masses given in eq.~(\ref{Eq:complicated}) valid in the alignment limit, we can eliminate combinations of masses and $\alpha_3$ appearing in (\ref{Eq:complicated}) to get
\bea \label{Eq:anotherequation}
&&-v^2\left[v_1^2(\Re\lambda_6)-v_2^2(\Re\lambda_7)\right]^2{\cal M}_{33}^2
+\frac{v^4v_1^2v_2^2}{v_1^4-v_2^4}(\Im \lambda_5)^2
\left(v_2^2{\cal M}_{11}^2-v_1^2{\cal M}_{22}^2\right)\\
&&-2v^3v_1{\cal M}_{13}^2\Im\lambda_5(v_1^2\Re\lambda_6-v_2^2\Re\lambda_7)\nonumber\\
&&
+\frac{2v^2v_2}{(v_1^4-v_2^4)v_1}\left[
v_1^2{\cal M}_{33}^2\left(v_2^2{\cal M}_{11}^2-v_1^2{\cal M}_{22}^2\right)
+(v_1^4-v_2^4)\left({\cal M}_{23}^2\right)^2
\right]
(v_1^2\Re\lambda_6+v_2^2\Re\lambda_7)=0.\nonumber
\eea
Here, the ${\cal M}_{ij}^2$ are elements of the mass-squared matrix in the neutral sector, directly related to the quartic couplings and vevs.
Together, these two relations (\ref{Eq:m12sqre}) and (\ref{Eq:anotherequation}) constitute the conditions for SCPV in the alignment limit. By substituting the expressions for the squared mass-matrix elements given in appendix~A of Ref.~\cite{Grzadkowski:2014ada} one can express them in terms of potential parameters and vevs only.
\subsubsection{Discussion of spontaneous CP violation in the alignment limit}
\label{Sec:SCPV-alignment-discussion}
Before closing this section let us collect here a few relevant comments:
\begin{itemize}
\item
If both alignment (conditions (\ref{eq:newalign1}) and (\ref{eq:newalign2})) and SCPV are imposed, then one finds by combining with Eq.~(\ref{Eq:m12sqre}) that
\beq
m_{12}^2=0 \lsp {\rm and} \lsp m_{11}^2=m_{22}^2.
\label{eq:newalign_scpv}
\eeq
It is worth noting  that the above conditions are basis independent -- if they are satisfied in one basis they hold in any basis.
\item 
Secondly, we notice that $\mu^2$ appears in (\ref{q1}) along with only masses of physical scalars. This means that $\mu^2$ itself must represent an observable quantity in the alignment limit for a basis in which $\xi=0$. It is instructive to allow here for non-zero $\xi$, then the coupling $q_1$ in the alignment limit becomes
\bea
q_{1}
&=&\frac{1}{v}\left(2 M_{H^\pm}^2-2\frac{\mu^2}{\cos\xi}+M_1^2\right),
\label{q1-xinonzero}
\eea
so in fact it is $\mu^2/\cos\xi$ that must be a basis-independent quantity in the alignment limit.
The freedom of choosing a different basis for $(\Phi_1,\Phi_2)$ can be parametrized by the relation
\beq \label{Eq:U2-rotation}
\left(
\begin{array}{c}\Phi_1\\ \Phi_2
\end{array}\right)
\rightarrow
\left(
\begin{array}{c}\bar{\Phi}_1\\ \bar{\Phi}_2
\end{array}\right)
=
e^{i\psi}\left(
\begin{array}{cc}\cos\theta & e^{-i\tilde\xi}\sin\theta\\ -e^{i\chi}\sin\theta & e^{i(\chi-\tilde\xi)}\cos\theta
\end{array}\right)
\left(
\begin{array}{c}\Phi_1\\ \Phi_2
\end{array}\right).
\eeq
It has obvious implications for the coefficients in the potential.
However, if the alignment conditions\footnote{
The conditions for alignment given in Eqs.~(\ref{eq:newalign1})	and (\ref{eq:newalign2}) for the $\xi=0$ basis will in a basis with non-zero $\xi$ be modified to	
\bea
\Im \left(e^{i\xi}m_{12}^2\right)&=&0,\\
m_{11}^2-m_{22}^2&=&\frac{\Re m_{12}^2}{\cos\xi}\left(\frac{v_1}{v_2}-\frac{v_2}{v_1}\right).
\eea
}
are imposed, then certain parameters (or combinations thereof) which in general (i.e., without the alignment) would be affected by the above unitary basis transformation remain unchanged. For instance, we find that
\beq
\frac{\bar{\mu}^2}{\cos\bar{\xi}} = \frac{\mu^2}{\cos\xi},
\label{mu_trans}
\eeq
under a general basis transformation with the restriction of alignment.
We conclude that this ratio is invariant under basis transformations, {\it provided we impose alignment}, hence in the alignment limit it is an observable quantity.

For other interesting properties of basis transformations see Appendix~\ref{Higgs_basis}.
\end{itemize}

\section{SCPV near alignment with softly broken \boldmath{$\zcal_2$} symmetry (``2HDM5'')}
\label{Sec:SCPV5}

As we have shown earlier \cite{Grzadkowski:2014ada}, {\it exact} alignment implies CP conservation in the case of  the 2HDM5. This is because within the 2HDM5 all the $\Im J_i$ vanish in the alignment limit.
One way to retain the possibility of having CP violation in the 2HDM5 setup is to allow for {\it small deviations} from alignment.
Then, when the alignment condition is relaxed, one can remain within the softly broken $\zcal_2$ symmetry 
without the necessity of introducing non-zero $\lambda_6$ and $\lambda_7$.
 
Here we adopt a basis such that $\lambda_6=0$ and $\lambda_7=0$ (no hard breaking of the $\zcal_2$ symmetry), however we allow for $m_{12}^2\neq 0$ 
(soft breaking is allowed). 

The 2HDM5 does not accommodate CP violation (neither explicit nor spontaneous) in the alignment limit. Therefore, we shall relax the alignment condition (\ref{Eq:alignment-angles})
by expanding our results around $\alpha_1=\beta$ and $\alpha_2=0$ in powers of $\delta\equiv \alpha_1-\beta$ and $\alpha_2$
for\footnote{We note that $\alpha_2=0$ implies CP conservation (independently of $\delta$). Thus, we require $\alpha_2\neq0$ for CP violation to occur.}
\begin{equation}
	|\delta|\ll1, \quad
	|\alpha_2|\ll1,
\end{equation}
and keeping leading terms in $\delta$ and $\alpha_2$. We find to lowest order :
\begin{equation}
	M_3^2-M_2^2=2 \alpha_2 \frac{M_2^2 - M_1^2}{\tan2\beta \sin2\alpha_3},
\end{equation}
and
\beq
\Im\lambda_5=\alpha_2\frac{M_2^2-M_1^2}{v^2c_\beta s_\beta}, \label{imlambda5}
\eeq
so that $\Im\lambda_5$ is non-zero as it needs to be in order to have CP violation in the 2HDM5.
The $I$-invariants are linear combinations of the $\Im J_i$-invariants as we see in Appendix~\ref{sect:results}. Expanding the $\Im J_i$ to the leading (linear) order in these small quantities we find:
\begin{align}
	\Im J_1&=\frac{\alpha_2 (M_2^2-M_1^2)(M_2^2-\mu^2)}{v^4c_\beta s_\beta}\, c_{2\beta}, \\
	\Im J_{11}&=\frac{\alpha_2 M_2^2(M_2^2-M_1^2)(M_2^2-\mu^2)}{v^6c_\beta s_\beta}\, c_{2\beta}
	=\frac{M_2^2}{v^2}\Im J_1,\\
	\Im J_2&=0,\\
	\Im J_{30}&=\frac{\alpha_2(M_2^2-M_1^2)(M_2^2-\mu^2)}{8v^6 c_\beta^3 s_\beta^3} \\
	&\times
	\bigl[-M_1^2-2(M_2^2+M_{H^\pm}^2-2\mu^2)+c_{4\beta}(M_1^2-2M_2^2+2M_{H^\pm}^2)
	\bigr]c_{2\beta} \nonumber \\
	&= \frac{1}{8v^2} \bigl[-M_1^2-2(M_2^2+M_{H^\pm}^2-2\mu^2)+c_{4\beta}(M_1^2-2M_2^2+2M_{H^\pm}^2)
	\bigr] \frac{\Im J_1}{c_\beta^2 s_\beta^2}.
\end{align}
So this explicitly shows that relaxing alignment allows for CP violation in the 2HDM5. In this "near-alignment" limit, we see that we have CP violation provided
\bea
M_2^2\neq M_1^2\quad \text{and}\quad \mu^2\neq M_2^2\quad \text{and}\quad \tan\beta\neq 1\, (v_1\neq v_2).
\eea
Next, let us study the conditions (\ref{eq:theorem-mch}) and (\ref{eq:theorem-q}) for spontaneous CP violation in order to figure out under which conditions the CP violation is in fact spontaneous. Remembering that it is the $I-invariants$ of  Appendix~\ref{sect:results} that ``control'' the CP invariance of the potential and that these are linear combinations of the $\Im J_i$, we get the leading (linear) order of the $I$-invariants by expanding the $\Im J_i$ to the linear order, while the coefficients (or prefactors) need only be expanded to the constant (leading) order. This means that the prefactors are identical in the near-alignment and the {\it exact} alignment limit. The conditions (\ref{eq:theorem-mch}) and (\ref{eq:theorem-q}) originate from prefactors only, meaning that they are represented by (\ref{Eq:muzero}) and (\ref{Eq:complicated}) also in the near-alignment limit. By putting $\lambda_6=\lambda_7=0$ and using (\ref{Eq:complicated}), we see that it is satisfied by default at the constant-order level. Thus, the only requirement for SCPV in the near-alignment limit becomes
\bea 
&&\mu^2=0.
\eea
We conclude that if CP is broken in the ``near alignment'' limit of the 2HDM5, then
$\mu^2=0$ guarantees that the CP violation is spontaneous. If $\mu^2\neq0$, the CP violation is explicit. The $\mu^2=0$ condition is also a necessary (but not sufficient condition) for SCPV in the alignment limit of the 2HDM7, as discussed in Section~\ref{Sec:SCPV-alignment-p67}.
\subsection{SCPV1 and SCPV2 in the near alignment limit of 2HDM5}
In ref.~\cite{Grzadkowski:2013rza} we have defined two possible scenarios, SCPV1 and SCPV2, for spontaneous CP violation in the 2HDM5. It is worth checking which of them could be realized in the ``near alignment'' limit.

\paragraph{SCPV1 near alignment.}

The condition for SCPV1 reads \cite{Grzadkowski:2013rza}
\beq
4\frac{\mu^2}{v^2}\Re\lambda_5-4\left(\frac{\mu^2}{v^2}\right)^2+(\Im\lambda_5)^2=0 \quad
({\rm or~equivalently} \quad
\Im\left[(m_{12}^2)^2\lambda_5^*\right]=0) \label{scpv1},
\eeq
provided we have CP violation \cite{Branco:1985aq}.
Near alignment the above condition reduces to
\begin{equation}
M_2^2\mu^2=0.
\end{equation}
Only $\mu^2=0$ is possible. This is consistent with the requirements for spontaneous CP violation from the previous section, and we conclude that SCPV1 is possible near the alignment region of the 2HDM5.

\paragraph{SCPV2 near alignment.}
The condition for SCPV2 reads \cite{Grzadkowski:2013rza}
\bea
\lambda_1&=&\lambda_2,\\
m_{11}^2&=&m_{22}^2,
\eea
(or equivalently $\lambda_1=\lambda_2$ and  $\lambda_1=\lambda_3+\lambda_4+\Re\lambda_5-2\mu^2/v^2$), again provided we have CP violation. The requirement for CP violation excludes the scenario where  $v_1= v_2$ ($\tan\beta=1$) since this would imply CP conservation (see CPC4 of \cite{Grzadkowski:2013rza}). In the ``near alignment'' limit, the above conditions reduce to
\bea
(M_2^2-\mu^2)\cos2\beta&=&0,\nonumber\\
\mu^2\cos 2\beta&=&0.
\eea

There is no solution satisfying both these two constraints that also allows for CP violation. Thus, SCPV2 is not possible in the ``near alignment'' region of the 2HDM5.

\section{Concluding remarks}
\label{Sect:conclusion}
We have seen that the conditions for spontaneous CP violation in the general 2HDM can be expressed in terms of physical quantities, and are remarkably simple. Apart from the masses of the three neutral Higgs bosons and their couplings to gauge bosons and to the charged pair, the charged-Higgs mass and its quartic coupling are involved. 

In \cite{Grzadkowski:2014ada}, we were able to express all the $\Im J_i$ in terms of masses and couplings, and also relate these invariants to physical processes which could allow for their measurement \cite{Grzadkowski:2016lpv}. It would be desirable to achieve the same for the two conditions (\ref{eq:theorem-mch}) and (\ref{eq:theorem-q}). However finding processes in which these expressions appear directly as part of the amplitude does not seem very easy. If we restrict ourselves to the alignment limit, (\ref{eq:theorem-mch}) and (\ref{eq:theorem-q}) simplify to Eqs.~(\ref{Eq:expression-couplings-alignment-1}) and (\ref{Eq:expression-couplings-alignment-2}).
Since we have not found measurements in which conditions (\ref{eq:theorem-mch}) and (\ref{eq:theorem-q}) appear explicitly, it is useful to consider two cases (in the alignment limit):
\bit
\item
In order to disprove SCPV a minimal set of measurements consists of $M_{H^\pm}$ and $q_1$, if they do not satisfy  
$q_1=\left(2M_{H^\pm}^2+M_1^2  \right)/v$, then CP is not violated spontaneously. 
\item 
To prove SCPV is strictly speaking impossible since one would need to show that 
equations (\ref{Eq:expression-couplings-alignment-1}) and (\ref{Eq:expression-couplings-alignment-2}) hold exactly. Since measurements are always subject to experimental (and theoretical) uncertainties, indeed, the above equations could at best only hold within some confidence level. Note, however, that the verification of the
above constraints require a determination of 9 parameters. $M_1$  and $v$ are already known, so 7 new measurements should be performed in order to test these constraints.
\item In the general case without alignment, we need to test the conditions (\ref{eq:theorem-mch}) and (\ref{eq:theorem-q}) in order to determine whether a measured CP violation is spontaneous or explicit.
In the general case without alignment, the 2HDM7 potential contains $14-3=11$ physical parameters\footnote{
	The potential for 2HDM7 contains 14 real parameters, 3 of which could be eliminated by a suitable choice of basis, therefore the
	number of physical parameters is indeed 11.}.
The above constraints also contain 11 independent masses and couplings altogether\footnote{Remembering that $v^2=e_1^2+e_2^2+e_3^2$.}.
Again, $M_1$  and $v$ are known, so 9 new measurements should be performed. Therefore we conclude that in order to test for
SCPV, all potential parameters must be known.
\eit

\section*{Acknowledgments}
It is a pleasure to thank H.~Haber and M. N. Rebelo for discussions.
BG acknowledges partial support by the National Science Centre (Poland) research project, 
decision no DEC-2014/13/B/ST2/03969. PO is supported in part by the Research Council of Norway.

\appendix
\section{Comments on the Higgs basis}
\label{Higgs_basis}
\subsection{The Higgs basis vs the {\boldmath $m_{12}^2=0$} basis}

The basis transformation (\ref{Eq:U2-rotation}) specified by  
\begin{subequations} \label{Eq:Higgsbasis}
\begin{align}
\tilde{\xi}&=\xi\equiv \xi_2-\xi_1,\\
\theta&=\beta=\arctan\left(\frac{v_2}{v_1}\right),\\
\chi&=0,\\
\psi&=-\xi_1.
\end{align}
\end{subequations}
leads to real vev's such that $\bar v_1=v$, $\bar\xi_1=0$ and $\bar v_2=0$ ($\bar\xi_2$ is undefined).
So this transformations brings us to the Higgs basis \cite{Donoghue:1978cj,Georgi:1978ri}.

The quadratic coefficient $m_{12}^2$ of the potential will under this rotation transform into
\bea
\bar{m}_{12}^2&=&\left[(-m_{11}^2+m_{22}^2)v_1v_2+\Re(m_{12}^2e^{i\xi})(v_1^2-v_2^2)+i\Im(m_{12}^2e^{i\xi})\right].
\eea
Enforcing the conditions of alignment given in (\ref{eq:newalign1}) and (\ref{eq:newalign2}), 
we find $\bar{m}_{12}^2=0$. Thus, the same transformation which makes the second vev vanish also eliminates the mixing parameter $m_{12}^2$ provided one remains in the alignment regime. 
As will be noticed below, $m_{12}^2=0$ is one the two alignment conditions in the Higgs basis,
which explains the above result.

Note also that since the definition of $\mu^2$, Eq.~(\ref{Eq:musq}), involves $v_1v_2$ in the denominator,
the rotation (\ref{Eq:Higgsbasis}) does not imply $\bar{\mu}^2=0$ since in the new basis $v_2$ vanishes.

\subsection{Alignment in the Higgs basis}

Using the most general conditions for the alignment
(\ref{align_lam_1}) and (\ref{align_lam_2}) it is easy to derive the conditions for alignment in the Higgs basis:
\beq \label{Eq:Higgs-basis-alignment}
m_{12}^2=0 \quad \text{and} \quad \lambda_6=0.
\eeq
In this basis one doublet couples to the vector bosons with full
strength, nevertheless it need not a mass eigenstate, to enforce  that one needs in addition to impose the above conditions (\ref{Eq:Higgs-basis-alignment}).

\section{Invariants for the most general case}
\label{sect:results}
\setcounter{equation}{0}

The four $I$ invariants can be expressed in reasonably compact form, in terms of masses and couplings, using the dimensionless quantities
\begin{align} \label{Eq:d_ijk}
	d_{ijk}&=\frac{q_{1}^i M_1^{2j}e_1^k+q_{2}^i M_2^{2j}e_2^k+q_{3}^i M_3^{2j}e_3^k}{v^{i+2j+k}},\\
	m_+&=\frac{M_{H^\pm}^2}{v^2}.
\end{align}
Below, we present them as expansions in the quantities $\Im J_1$, $\Im J_{11}$, $\Im J_2$ and $\Im J_{30}$.
At least one of these must be non-zero for there to be any CP violation at all.
We note that while this is a convenient base, it is over-complete, $\Im J_{11}$ vanishes whenever $\Im J_1$ and $\Im J_2$ both vanish.

\subsection{The invariant \boldmath{$I_{Y3Z}$}}
\bea
\frac{I_{Y3Z}}{v^2}
&=&
\left(d_{010} d_{012}-d_{010} d_{101}+2 d_{010} m_+-2 d_{012} m_+-d_{022}-2 d_{101} m_++d_{200}\right)\Im J_1\nonumber\\
&&
+\left(-d_{012}+2 d_{101}-4 m_+-2 q\right)\Im J_{11}
+\left(\frac{d_{101}}{2}-m_+-q\right)\Im J_2\nonumber\\
&&
+\left(-d_{012}+d_{101}-2 m_+\right)\Im J_{30}\label{eq:IY3Z}
\eea
\subsection{The invariant \boldmath{$I_{2Y2Z}$}}
\bea
\frac{I_{2Y2Z}}{v^4}
&=&\left(\frac{d_{010} d_{012}}{2}-\frac{d_{010} d_{101}}{2}+d_{010} m_+-d_{012} m_+-\frac{d_{022}}{2}+\frac{d_{111}}{2}\right)\Im J_1\nonumber\\
&&
+\left(-\frac{d_{012}}{2}+\frac{d_{101}}{2}-m_+\right)\Im J_{11}
+\left(\frac{d_{101}}{4}-\frac{m_+}{2}-\frac{q}{2}\right)\Im J_2\label{eq:I2Y2Z}
\eea
\subsection{The invariant \boldmath{$I_{6Z}$}}
\bea
I_{6Z}&=&
\left(4 d_{010}^3 d_{012} -8 d_{010}^2 d_{101}^2 +16 d_{010}^2 q^2 -12
  d_{010}^2 d_{022} +8 d_{010}^2 d_{200} -8 d_{010}^2 d_{012} q \right.\nonumber\\
&&\hspace*{0.2cm}
-8 d_{010}^2 d_{012} m_+ +16 d_{010}^2q m_+ +4 d_{010} d_{101}^3 +2 d_{010} d_{012} d_{101}^2 -40 d_{010} d_{012} q^2 \nonumber\\
&&\hspace*{0.2cm}+8 d_{010} d_{101} q^2 +8 d_{010} d_{012} m_+^2 -16 d_{010} q m_+^2 -4 d_{010} d_{012} d_{020} +4 d_{010} d_{012} d_{022} \nonumber\\
&&\hspace*{0.2cm}+12 d_{010} d_{032} -2 d_{010} d_{012}^2 d_{101} +4 d_{010} d_{022} d_{101} +36  d_{010} d_{101} d_{111}-16 d_{010} d_{012} d_{200} \nonumber\\
&&\hspace*{0.2cm}-4  d_{010} d_{101} d_{200} -24 d_{010} d_{210}-4 d_{010} d_{012}^2 q +12 d_{010} d_{101}^2 q -16 d_{010} d_{020} q \nonumber\\
&&\hspace*{0.2cm}+40 d_{010} d_{022} q -20 d_{010} d_{012} d_{101} q +12 d_{010} d_{111} q -8 d_{010} d_{200} q +8 d_{010} d_{101}^2 m_+ \nonumber\\
&&\hspace*{0.2cm}-16 d_{010} q^2 m_+ +20 d_{010} d_{022} m_+ -8 d_{010} d_{200} m_+ -32 d_{010} d_{012} q m_+ -4 d_{022}^2-4 d_{200}^2\nonumber\\
&&\hspace*{0.2cm}-16 d_{020} q^2+40 d_{022} q^2-8 d_{111} q^2-8 d_{012}^2 m_+^2-8 d_{012} d_{101} m_+^2+16 d_{012} q m_+^2\nonumber\\
&&\hspace*{0.2cm}+16 d_{101} q m_+^2-4 d_{012} d_{020} d_{101}+2 d_{012} d_{022} d_{101}+6 d_{012}^2 d_{111}-8 d_{101}^2 d_{111}+8 d_{020} d_{111}\nonumber\\
&&\hspace*{0.2cm}-16 d_{022} d_{111}-4 d_{012} d_{101} d_{111}+2 d_{012}^2 d_{200}+4 d_{101}^2 d_{200}+8 d_{022} d_{200}+6 d_{012} d_{101} d_{200}\nonumber\\
&&\hspace*{0.2cm}+32 d_{012} d_{020} q+4 d_{012} d_{022} q+16 d_{030} q-64 d_{032} q-8 d_{020} d_{101} q+24 d_{022} d_{101} q-4 d_{012} d_{111} q\nonumber\\
&&\hspace*{0.2cm}-16 d_{101} d_{111} q+4 d_{012} d_{200} q+4 d_{101} d_{200} q-8 d_{101}^3 m_+-4 d_{012} d_{101}^2 m_++16 d_{012} q^2 m_+\nonumber\\
&&\hspace*{0.2cm}-12 d_{032} m_+-8 d_{022} d_{101} m_+-8 d_{101} d_{111} m_++12 d_{012} d_{200} m_++8 d_{101} d_{200} m_++8 d_{210} m_+\nonumber\\
&&\left.\hspace*{0.2cm}+8 d_{012}^2 q m_+-8 d_{101}^2 q m_++8 d_{022} q m_++16 d_{012} d_{101} q m_+-8 d_{200} q m_+\right)\Im J_1\nonumber\\
&&+\left(-8 d_{010}^2 d_{012} +16 d_{010}^2 q-16 d_{010} q^2+20 d_{010} d_{022} -8  d_{010} d_{111}-32 d_{010} d_{012} q +16 d_{010} d_{101} q \right.\nonumber\\
&&\hspace*{0.6cm}+16 d_{010} d_{012} m_+ -32 d_{010} q m_+ +32 d_{012} q^2-16 d_{101} q^2-16 d_{012} m_+^2+32 q m_+^2+8 d_{012} d_{020}\nonumber\\
&&\hspace*{0.6cm}-4 d_{012} d_{022}-16 d_{032}-4 d_{012}^2 d_{101}+8 d_{022} d_{101}+8 d_{012} d_{111}-40 d_{101} d_{111}+20 d_{012} d_{200}\nonumber\\
&&\hspace*{0.6cm}+16 d_{210}-16 d_{101}^2 q-16 d_{020} q+24 d_{022} q+8 d_{012} d_{101} q-24 d_{111} q+24 d_{200} q+4 d_{012}^2 m_+\nonumber\\
&&\hspace*{0.6cm}-8 d_{101}^2 m_++32 q^2 m_+-28 d_{022} m_++16 d_{012} d_{101} m_++8 d_{200} m_+\nonumber\\
&&\left.\hspace*{0.6cm}+32 d_{012} q m_+-32 d_{101} q m_+\right)\Im J_{11}\nonumber\\
&&+\left(8 q^3-8 d_{010} q^2+4 d_{012} q^2-8 d_{101} q^2+16 q^2 m_+ +4 d_{010}^2 q-8 d_{101}^2 q+8 q m_+^2 -4 d_{010} d_{012} q\right.\nonumber\\
&&\hspace*{0.6cm}-4 d_{020} q+4 d_{022} q+4 d_{010} d_{101} q+10 d_{200} q-8 d_{010} q m_+ -8 d_{101} q m_+ -d_{012} d_{101}^2\nonumber\\
&&\hspace*{0.6cm}-4 d_{012} m_+^2-2 d_{010}^2 d_{012}+2 d_{012} d_{020}+4 d_{010} d_{022}-4 d_{032}+2 d_{022} d_{101}-2 d_{010} d_{111}\nonumber\\
&&\left.\hspace*{0.6cm}-4 d_{101}^2 m_++4 d_{010} d_{012} m_+-4 d_{022} m_+-2 d_{012} d_{101} m_++6 d_{111} m_++4 d_{200} m_+\right)\Im J_2\nonumber\\
&&
+\left(6 d_{012}^3+4 d_{010} d_{012}^2-8  d_{012}^2 m_+ -4 d_{010}^2 d_{012}+2 d_{012} d_{101}^2 -8 d_{012} m_+^2 +4 d_{012} d_{020} \right.\nonumber\\
&&\hspace*{0.6cm}-20 d_{012} d_{022}+12 d_{010} d_{012} d_{101} +12  d_{012} d_{200}-24 d_{010} d_{012} q +8 d_{010} d_{012} m_+ \nonumber\\
&&\hspace*{0.6cm}+32 d_{012} q m_+ +4 d_{101}^3+8 d_{010} d_{101}^2+16 q m_+^2+8 d_{010} d_{022}+8 d_{022} d_{101}-16 d_{010} d_{111}\nonumber\\
&&\hspace*{0.6cm}-24 d_{101} d_{111}-8 d_{010} d_{200}-4 d_{101} d_{200}+8 d_{210}+8 d_{010}^2 q+4 d_{101}^2 q-8 d_{020} q\nonumber\\
&&\hspace*{0.6cm}+28 d_{022} q+8 d_{010} d_{101} q-16 d_{111} q-8 d_{101}^2 m_+-8 d_{022} m_++8 d_{111} m_+\nonumber\\
&&\left.\hspace*{0.6cm}+8 d_{200} m_+-16 d_{010} q m_+-16 d_{101} q m_+\right)\Im J_{30}\nonumber\\\label{eq:I6Z}
\eea

\subsection{The invariant \boldmath{$I_{3Y3Z}$}}
\bea
\frac{I_{3Y3Z}}{v^6}&=&
\left(\frac{d_{010}^3 d_{012}}{4}-\frac{3 d_{010}^2 d_{022} }{4}+d_{010} d_{012} m_+^2 +\frac{5  d_{010} d_{012} d_{022}}{4}+\frac{ d_{010} d_{032}}{4}+\frac{1}{4} d_{010} d_{012}^2 d_{101} \right.\nonumber\\
&&\hspace*{0.6cm}+\frac{ d_{010} d_{101} d_{111}}{4}-d_{010} d_{012}^2 m_+ +\frac{3}{2} d_{010} d_{022} m_+ -d_{010} d_{111} m_+ -\frac{3  d_{010} d_{012} d_{020}}{4}\nonumber\\
&&\hspace*{0.6cm}-\frac{d_{010} d_{022} d_{101} }{4}-\frac{d_{010} d_{012} d_{111} }{4}+\frac{d_{022} d_{101}^2}{2}-2 d_{012}^2 m_+^2-d_{020} m_+^2+2 d_{022} m_+^2\nonumber\\
&&\hspace*{0.6cm}-d_{012} d_{101} m_+^2+\frac{d_{020} d_{022}}{2}+\frac{d_{012} d_{030}}{2}+\frac{d_{012} d_{032}}{4}+\frac{d_{012} d_{020} d_{101}}{2}+\frac{d_{101}^2 d_{111}}{4}\nonumber\\
&&\hspace*{0.6cm}+\frac{d_{022} d_{111}}{2}+\frac{d_{012}^3 m_+}{2}+\frac{1}{2} d_{012} d_{101}^2 m_+-d_{012} d_{020} m_++d_{012}^2 d_{101} m_++d_{020} d_{101} m_+\nonumber\\
&&\hspace*{0.6cm}-\frac{5}{2} d_{022} d_{101} m_++d_{012} d_{111} m_+-d_{101} d_{111} m_++\frac{1}{2} d_{012} d_{200} m_++d_{210} m_+-\frac{3 d_{022}^2}{2}\nonumber\\
&&\left.\hspace*{0.6cm}-\frac{d_{032} d_{101}}{2}-\frac{d_{020} d_{101}^2}{4}-\frac{d_{012}^2 d_{111}}{4}-\frac{d_{012} d_{101} d_{111}}{4}-\frac{d_{012} d_{101} d_{200}}{4}-\frac{d_{101} d_{210}}{4}\right)\Im J_1\nonumber\\
&&+\left(-\frac{d_{101}^3}{4}+\frac{d_{010} d_{101}^2}{4}+ d_{101}^2 m_+-\frac{d_{012} d_{101}^2}{4}+\frac{d_{022} d_{101}}{2}+\frac{d_{101} d_{200} }{4}+\frac{d_{012} d_{101} q }{2}\right.\nonumber\\
&&\hspace*{0.6cm}-d_{010} d_{101} m_+ +d_{012} d_{101} m_+ -\frac{d_{010} d_{012} d_{101}}{2}-\frac{d_{101} d_{111} }{2}-\frac{d_{012}^2 d_{101}}{4}+d_{010} m_+^2\nonumber\\
&&\hspace*{0.6cm}-3 d_{012} m_+^2+\frac{d_{010} d_{022}}{4}+d_{032}+\frac{d_{012} d_{111}}{2}+\frac{d_{012}^2 m_+}{2}+d_{010} d_{012} m_+-2 d_{022} m_+\nonumber\\
&&\left.\hspace*{0.6cm}+2 d_{111} m_+-d_{200} m_+-d_{012} q m_+-\frac{5 d_{012} d_{022}}{4}\right)\Im J_{11}\nonumber\\
&&+\left(-\frac{d_{101}^3}{4}+\frac{d_{010} d_{101}^2}{8}+\frac{3 d_{101}^2 m_+}{4}+\frac{d_{010}^2 d_{101}}{4}+\frac{3 d_{022} d_{101}}{4}+\frac{d_{101} d_{200} }{4}+\frac{d_{012} d_{101} q }{2}\right.\nonumber\\
&&\hspace*{0.6cm}-\frac{1}{2} d_{010} d_{101} m_+ +\frac{1}{2} d_{012} d_{101} m_+ -\frac{d_{020} d_{101}}{4}-\frac{d_{010} d_{101} q}{4}+\frac{d_{010} m_+^2}{2}-d_{012} m_+^2\nonumber\\
&&\hspace*{0.6cm}+\frac{d_{012} d_{020}}{4}+\frac{d_{010} d_{022}}{2}+\frac{d_{012}^2 q}{4}+\frac{d_{010} d_{012} q}{4}+\frac{d_{012}^2 m_+}{4}+\frac{5}{4} d_{010} d_{012} m_++\frac{d_{020} m_+}{2}\nonumber\\
&&\hspace*{0.6cm}+\frac{d_{111} m_+}{2}+\frac{1}{2} d_{010} q m_+-d_{012} q m_+-\frac{d_{032}}{2}-\frac{d_{022} q}{2}-\frac{d_{010}^2 m_+}{2}-\frac{3 d_{022} m_+}{2}\nonumber\\
&&\left.\hspace*{0.6cm}-\frac{d_{010}^2 d_{012}}{4}-\frac{d_{012} d_{200}}{4}-\frac{3 d_{200} m_+}{4}-\frac{5 d_{010} d_{111}}{8}-\frac{d_{012} d_{111}}{8}\right)\Im J_2\nonumber\\
&&
+\left(\frac{d_{012}^2 d_{101} }{4}- d_{012}m_+^2+d_{111} m_+-\frac{d_{022} m_+}{2}-\frac{d_{101} d_{111}}{4}\right)\Im J_{30}\label{eq:I3Y3Z}
\eea
\subsection{Case Study: \boldmath{$I_{Y3Z}$}}
We shall briefly outline how we were able to express the $I$-invariants in terms of masses and couplings, by using $I_{Y3Z}$ as an example.
We know that all the $I$-invariants should vanish in the CP conserving limit (all $\Im J_i=0$), so we start with the ansatz\footnote{The ansatz $I_{Y3Z}=v^2\left(C_1 \Im J_1 + C_2 \Im J_2 + C_{30}\Im J_{30}\right)$ led nowhere. This changed after inclusion of the $\Im J_{11}$ term. We recall from \cite{Grzadkowski:2014ada} that $\Im J_1=\Im J_2=0 \implies \Im J_{11}=0$.} that 
\bea
I_{Y3Z}&=&v^2\left(C_1 \Im J_1 +C_{11} \Im J_{11} + C_2 \Im J_2 + C_{30}\Im J_{30}\right)\label{eq:ansatz}.
\eea
In \cite{Grzadkowski:2014ada}, all the $\Im J_i$ were expressed in terms of masses and couplings. These expression were all antisymmetric under exchange of two of the indices $\{i,j,k\}=\{1,2,3\}$. Assuming that this is true for all CP violating invariants, we expect this to be the case also for $I_{Y3Z}$. This implies that all the coefficients $C_i$ should be symmetric under exchange of two of the indices $\{i,j,k\}=\{1,2,3\}$ when expressed in terms of masses and couplings.

Next, we note that $I_{Y3Z}$ is a homogeneous polynomial of order 4 in the variables of the set ${\cal P}_0$ (see Ref.~\cite{Grzadkowski:2014ada}). Our ansatz then implies that the right hand side of (\ref{eq:ansatz}) must also be of order 4. We know that $\Im J_{1}$ is of order 2, whereas $\Im J_{11}, \Im J_2$ and $\Im J_{30}$ are of order 3. This implies the order of the coefficients $C_i$. $C_{1}$ must be of order 2, while $C_{11}, C_2$ and $C_{30}$ must be of order 1. All $C_i$ must be dimensionless, which motivates us to introduce the dimensionless and symmetric quantities $q$, $m_+$ and $d_{ijk}$, which we use to construct general symmetric expressions for the $C_i$ of the correct order. We continue with the ansatz that the general forms of the $C_i$ can be written as
\bea
C_1&=&c_{1}q^2+c_{2}qm_++c_{3}m_+^2
+ c_{4}qd_{010}+c_{5}qd_{012}+c_{6}qd_{101}\nonumber\\
&&+ c_{7}m_+d_{010}+c_{8}m_+d_{012}+c_{9}m_+d_{101}\nonumber\\
&&+c_{10}d_{010}^2+c_{11}d_{010}d_{012}+c_{12}d_{010}d_{101}+c_{13}d_{012}^2+c_{14}d_{012}d_{101}+c_{15}d_{101}^2\nonumber\\
&&+c_{16}d_{020}+c_{17}d_{022}+c_{18}d_{200}+c_{19}d_{111},\\
C_{11}&=&c_{20}q+c_{112}m_+ + c_{21}d_{010}+c_{22}d_{012}+c_{23}d_{101},\\
C_{2}&=&c_{24}q+c_{25}m_+ + c_{26}d_{010}+c_{27}d_{012}+c_{28}d_{101},\\
C_{30}&=&c_{29}q+c_{30}m_+ + c_{31}d_{010}+c_{32}d_{012}+c_{33}d_{101},
\eea
where the numerical coefficients $c_i$ are to be determined. In order to determine these, we expand each side of (\ref{eq:ansatz}) in the variables of ${\cal P}_0$, equating the coefficients of each such term. This makes it possible to determine all $c_i$, and the result found for $I_{Y3Z}$ can be read off (\ref{eq:IY3Z}).

The process is similar for $I_{2Y2Z}$, resulting in a different set of coefficients $c_i$. For $I_{6Z}$ and $I_{3Y3Z}$ all the coefficients $C_i$ are of order two higher than for $I_{Y3Z}$, so their general form is more complicated, but apart from this, the process is similar.

\section{Disentangling the \boldmath{$I$}-invariants}
\label{sect:newresults}
\setcounter{equation}{0}
Since all four invariants (\ref{eq:IY3Z}), (\ref{eq:I2Y2Z}), (\ref{eq:I6Z}) and (\ref{eq:I3Y3Z}) must vanish for the CP violation to be spontaneous, let us start by discussing the two simpler ones, $I_{Y3Z}$ and $I_{2Y2Z}$. These are both {\it linear} in $q$ and $M_{H^\pm}^2$. We put these expressions equal to zero, and treat the resulting two equations as a system of two linear equations with two unknowns ($q$ and $M_{H^\pm}^2$). Whenever this system is non-singular, we can solve it uniquely for the two unknowns. In order to determine when the system is singular, we calculate the determinant of the coefficient matrix, which (ignoring constant factors and powers of $v$) is found to be 
\bea
\Delta\propto D(\Im J_1)^2
\eea
with 
\begin{equation}
	D=e_1^2M_2^2M_3^2+e_2^2M_3^2M_1^2+e_3^2 M_1^2 M_2^2\,.
\end{equation}
For a physical system (positive $M_i^2$ and $e_1^2+e_2^2+e_3^2=v^2>0$), $D$ is positive definite. We conclude that the determinant vanishes if and only if $\Im J_1$ vanishes. We identify four different cases which we need to study separately.
\begin{alignat}{2}
	&\!\!\bullet &\ \
	&\text{\bf Case 0: } \Im J_1\neq 0 \nonumber \\
	&\!\!\bullet &\ \
	&\text{\bf Case 1: } \Im J_1= 0 \text{ because } e_i=v, e_j=0, e_k=0 \nonumber \\
	&\!\!\bullet &\ \
	&\text{\bf Case 2: } \Im J_1= 0 \text{ because } M_j^2=M_i^2, e_k=0 \nonumber \\
	&\!\!\bullet &\ \
	&\text{\bf Case 3: } \Im J_1= 0 \text{ because } q_k=\frac{e_je_kq_i(M_k^2-M_j^2)+e_ie_kq_j(M_i^2-M_k^2)}{e_ie_j(M_i^2-M_j^2)} \nonumber
\end{alignat}
Some comments are here in order. In cases 1-3 we do not include those scenarios where $\Im J_1=0$ which leads to CP conservation (scenarios where all $\Im J_i=0$), since then we cannot have spontaneous CP violation. All such cases are listed as six bullet points in Section 3.2 of our previous work \cite{Grzadkowski:2014ada}. Case 3 are those scenarios where $\Im J_1=0$ can be solved for one of the $q_k$. Cases 1 and 2 covers those scenarios where $\Im J_1=0$ but we cannot solve for any $q_k$.\\

{\bf Case~0~($\Im J_1\neq 0$):}\\
Solving the system of two equations for $q$ and $M_{H^\pm}^2$ we find
\begin{align} \label{Eq:Mh_ch_condition}
	M_{H^\pm}^2&=\frac{v^2}{2D}
	[e_1q_1M_2^2M_3^2 + e_2q_2 M_3^2 M_1^2 + e_3 q_3 M_1^2 M_2^2-M_1^2 M_2^2 M_3^2], \\
	q&=\frac{1}{2D}
	[(e_2q_3-e_3q_2)^2M_1^2+(e_3q_1-e_1q_3)^2M_2^2+(e_1q_2-e_2q_1)^2M_3^2 + M_1^2 M_2^2 M_3^2]
	\label{Eq:q_condition}.
\end{align}
Substituting these expressions for $q$ and $M_{H^\pm}^2$ into the remaining invariants $I_{6Z}$ and $I_{3Y3Z}$ we find that they both vanish, so the conditions for SCPV are in this case simply given by (\ref{Eq:Mh_ch_condition}) and (\ref{Eq:q_condition}).
\\

{\bf Case~1~($ e_i=v, e_j=0, e_k=0 $):}\\
In these cases we find that $I_{2Y2Z}=I_{3Y3Z}=0$. We solve $I_{Y3Z}=0$ for $M_{H^\pm}^2$ and substitute our expression for $M_{H^\pm}^2$ into $I_{6Z}=0$ which now is linear in $q$. This means that we may also solve for $q$. Considering for illustration $e_1=v, e_2=e_3=0$ (the alignment limit), we arrive at the following expressions for $q$ and $M_{H^\pm}^2$
\bea
M_{H^\pm}^2&=&\frac{vq_1-M_1^2}{2},\label{Eq:Mh_ch_condition-case1}\\
q&=&\frac{1}{2}\left(\frac{q_2^2}{M_2^2}+\frac{q_3^2}{M_3^2}+\frac{M_1^2}{v^2}\right).\label{Eq:q_condition-case1}
\eea
The other sub-cases are obtained by a cyclic rotation of $\{i,j,k\}=\{1,2,3\}$.
Comparing with (\ref{Eq:Mh_ch_condition}) and (\ref{Eq:q_condition}), we see that by simply putting $e_1=v, e_2=e_3=0$ into (\ref{Eq:Mh_ch_condition}) and (\ref{Eq:q_condition}), we arrive at (\ref{Eq:Mh_ch_condition-case1}) and (\ref{Eq:q_condition-case1}). \\

{\bf Case~2~($ M_j^2=M_i^2, e_k=0$):}\\
Also in these cases we find that $I_{2Y2Z}=I_{3Y3Z}=0$. We proceed as in the previous case, solving $I_{Y3Z}=0$ for $M_{H^\pm}^2$ and substitute our expression for $M_{H^\pm}^2$ into $I_{6Z}=0$ which now is linear in $q$. Considering for illustration $M_2^2=M_1^2, e_3=0$, we arrive at the following expressions for $q$ and $M_{H^\pm}^2$
\bea
M_{H^\pm}^2&=&\frac{e_1 q_1 + e_2 q_2 - M_1^2}{2},\label{Eq:Mh_ch_condition-case2}\\
q&=&\frac{v^2q_3^2M_1^2+(e_2q_1-e_1q_2)^2M_3^2+M_1^4M_3^2}{2v^2M_1^2M_3^2}\label{Eq:q_condition-case2}.
\eea
The other sub-cases are obtained by a cyclic rotation of $\{i,j,k\}=\{1,2,3\}$.
Comparing with (\ref{Eq:Mh_ch_condition}) and (\ref{Eq:q_condition}), we see that by simply putting $M_2^2=M_1^2, e_3=0$ into (\ref{Eq:Mh_ch_condition}) and (\ref{Eq:q_condition}), we arrive at (\ref{Eq:Mh_ch_condition-case2}) and (\ref{Eq:q_condition-case2}). \\

{\bf Case~3:}\\
These cover the cases where all $e_i\neq 0$ and all masses are non-degenerate. For illustration, we solve $\Im J_i=0$ for $q_3$ to get
\bea \label{Eq:case3:q_3}
q_3=\frac{e_2e_3q_1(M_3^2-M_2^2)+e_1e_3q_2(M_1^2-M_3^2)}{e_1e_2(M_1^2-M_2^2)}.
\eea
In this case we find that both $I_{Y3Z}$ and $I_{2Y2Z}$ contain the same factor which needs to vanish in order for the invariants to vanish. This factor is linear in both $q$ and $M_{H^\pm}^2$. We choose to solve for $q$ and then insert the expression for $q$ into $I_{6Z}=0$ and $I_{3Y3Z}=0$. These two equations also contains the same factor which needs to vanish in order for the invariants to vanish. The factor is now is linear in $M_{H^\pm}^2$, so we may solve for $M_{H^\pm}^2$. The result is substituted into the expression we found for $q$, and we arrive at
\begin{align}
	M_{H^\pm}^2&=
	\frac{v^2 e_3^2M_1^2M_2^2}{2e_1e_2(M_1^2-M_2^2)D}
	\bigl[e_2q_1(M_3^2-M_2^2)+e_1q_2(M_1^2-M_3^2)\bigr] \nonumber \\
	&+\frac{v^2M_3^2}{2D}
	[e_1q_1M_2^2+e_2q_2M_1^2-M_1^2M_2^2],\label{Eq:Mh_ch_condition-case3}\\
	q&=\frac{M_1^2M_2^2M_3^2}{2D}
	+\frac{(e_1q_2-e_2q_1)^2}{2e_1^2e_2^2(M_1^2-M_2^2)^2D} \nonumber \\
	&\times[e_1^2e_2^2M_3^2(M_1^2-M_2^2)^2
	+e_2^2e_3^2M_1^2(M_2^2-M_3^2)^2
	+e_3^2e_1^2M_2^2(M_3^2-M_1^2)^2].\label{Eq:q_condition-case3}
\end{align}
The other sub-cases are obtained by a cyclic rotation of $\{i,j,k\}=\{1,2,3\}$.
Comparing with (\ref{Eq:Mh_ch_condition}) and (\ref{Eq:q_condition}), we see that by simply putting the value for $q_3$ from Eq.~(\ref{Eq:case3:q_3}) into (\ref{Eq:Mh_ch_condition}) and (\ref{Eq:q_condition}), we arrive at (\ref{Eq:Mh_ch_condition-case3}) and (\ref{Eq:q_condition-case3}).



\begin{thebibliography}{99}

\bibitem{Lavoura:1994fv}
  L.~Lavoura and J.~P.~Silva,
  {\it Fundamental CP violating quantities in a SU(2) x U(1) model with many Higgs doublets,}
  Phys.\ Rev.\ D {\bf 50} (1994) 4619
  doi:10.1103/PhysRevD.50.4619
  [hep-ph/9404276].
    
\bibitem{Botella:1994cs}
  F.~J.~Botella and J.~P.~Silva,
  {\it Jarlskog-like invariants for theories with scalars and fermions,}
  Phys.\ Rev.\ D {\bf 51} (1995) 3870
  doi:10.1103/PhysRevD.51.3870
  [hep-ph/9411288].
  
\bibitem{Grzadkowski:2014ada}
  B.~Grzadkowski, O.~M.~Ogreid and P.~Osland,
  {\it Measuring CP violation in Two-Higgs-Doublet models in light of the LHC Higgs data,}
  JHEP {\bf 1411} (2014) 084
  doi:10.1007/JHEP11(2014)084
  [arXiv:1409.7265 [hep-ph]].

\bibitem{Branco:2005em}
  G.~C.~Branco, M.~N.~Rebelo and J.~I.~Silva-Marcos,
  {\it CP-odd invariants in models with several Higgs doublets,}
  Phys.\ Lett.\ B {\bf 614} (2005) 187
  doi:10.1016/j.physletb.2005.03.075
  [hep-ph/0502118].
  
\bibitem{Gunion:2005ja}
  J.~F.~Gunion and H.~E.~Haber,
  {\it Conditions for CP-violation in the general two-Higgs-doublet model,}
  Phys.\ Rev.\ D {\bf 72} (2005) 095002
  doi:10.1103/PhysRevD.72.095002
  [hep-ph/0506227].
  
\bibitem{Asner:2013psa}
  D.~M.~Asner {\it et al.},
  {\it ILC Higgs White Paper,}
  arXiv:1310.0763 [hep-ph].
  
\bibitem{Craig:2013hca}
  N.~Craig, J.~Galloway and S.~Thomas,
  {\it Searching for Signs of the Second Higgs Doublet,}
  arXiv:1305.2424 [hep-ph].
  
\bibitem{Carena:2013ooa}
  M.~Carena, I.~Low, N.~R.~Shah and C.~E.~M.~Wagner,
  {\it Impersonating the Standard Model Higgs Boson: Alignment without Decoupling,}
  JHEP {\bf 1404} (2014) 015
  doi:10.1007/JHEP04(2014)015
  [arXiv:1310.2248 [hep-ph]].

\bibitem{Branco:1999fs}
  G.~C.~Branco, L.~Lavoura and J.~P.~Silva,
  {\it CP Violation,}
  Int.\ Ser.\ Monogr.\ Phys.\  {\bf 103} (1999) 1.
  
\bibitem{Accomando:2006ga}
  E.~Accomando {\it et al.},
  {\it Workshop on CP Studies and Non-Standard Higgs Physics,}
  hep-ph/0608079.

\bibitem{ElKaffas:2006gdt}
  A.~W.~El Kaffas, W.~Khater, O.~M.~Ogreid and P.~Osland,
  {\it Consistency of the two Higgs doublet model and CP violation in top production at the LHC,}
  Nucl.\ Phys.\ B {\bf 775} (2007) 45
  doi:10.1016/j.nuclphysb.2007.03.041
  [hep-ph/0605142].

\bibitem{Davidson:2005cw}
  S.~Davidson and H.~E.~Haber,
  {\it Basis-independent methods for the two-Higgs-doublet model,}
  Phys.\ Rev.\ D {\bf 72} (2005) 035004
   Erratum: [Phys.\ Rev.\ D {\bf 72} (2005) 099902]
  doi:10.1103/PhysRevD.72.099902, 10.1103/PhysRevD.72.035004
  [hep-ph/0504050].

\bibitem{Maniatis:2007vn}
  M.~Maniatis, A.~von Manteuffel and O.~Nachtmann,
  Eur.\ Phys.\ J.\ C {\bf 57} (2008) 719
  doi:10.1140/epjc/s10052-008-0712-5
  [arXiv:0707.3344 [hep-ph]].

\bibitem{Khater:2003wq}
  W.~Khater and P.~Osland,
  {\it CP violation in top quark production at the LHC and two Higgs doublet models,}
  Nucl.\ Phys.\ B {\bf 661} (2003) 209
  doi:10.1016/S0550-3213(03)00300-6
  [hep-ph/0302004].
	
\bibitem{Dev:2014yca} 
  P.~S.~Bhupal Dev and A.~Pilaftsis,
  {\it Maximally Symmetric Two Higgs Doublet Model with Natural Standard Model Alignment,}
  JHEP {\bf 1412}, 024 (2014)
  Erratum: [JHEP {\bf 1511}, 147 (2015)]
  doi:10.1007/JHEP11(2015)147, 10.1007/JHEP12(2014)024
  [arXiv:1408.3405 [hep-ph]].
 
\bibitem{Grzadkowski:2013rza}
  B.~Grzadkowski, O.~M.~Ogreid and P.~Osland,
  {\it Diagnosing CP properties of the 2HDM,}
  JHEP {\bf 1401} (2014) 105
  doi:10.1007/JHEP01(2014)105
  [arXiv:1309.6229 [hep-ph]].

\bibitem{Branco:1985aq}
  G.~C.~Branco and M.~N.~Rebelo,
  {\it The Higgs Mass in a Model With Two Scalar Doublets and Spontaneous {CP} Violation,}
  Phys.\ Lett.\ B {\bf 160} (1985) 117.
  doi:10.1016/0370-2693(85)91476-5
  
\bibitem{Grzadkowski:2016lpv}
  B.~Grzadkowski, O.~M.~Ogreid and P.~Osland,
  {\it CP-Violation in the $ZZZ$ and $ZWW$ vertices at $e^+e^-$ colliders in Two-Higgs-Doublet Models,}
  JHEP {\bf 1605} (2016) 025
  doi:10.1007/JHEP05(2016)025
  [arXiv:1603.01388 [hep-ph]].

\bibitem{Donoghue:1978cj}
  J.~F.~Donoghue and L.~F.~Li,
  {\it Properties of Charged Higgs Bosons,}
  Phys.\ Rev.\ D {\bf 19} (1979) 945.
  doi:10.1103/PhysRevD.19.945

\bibitem{Georgi:1978ri}
  H.~Georgi and D.~V.~Nanopoulos,
  {\it Suppression of Flavor Changing Effects From Neutral Spinless Meson Exchange in Gauge Theories,}
  Phys.\ Lett.\ B {\bf 82} (1979) 95.
  doi:10.1016/0370-2693(79)90433-7
   
\end{thebibliography}
\end{document}